\documentclass[reprint,twocolumn,showpacs,preprintnumbers,amsmath,amssymb,superscriptaddress]{revtex4}

\usepackage{graphicx}
\usepackage{dcolumn}
\usepackage{bm}
\usepackage{color}

\begin{document}

\setlength{\footnotesep}{0.2cm}

\title{Fragmentation and systematics of the Pygmy Dipole Resonance in the stable N=82 isotones}

\author{D. Savran}%
\email{d.savran@gsi.de}
\affiliation{ExtreMe Matter Institute EMMI and Research Devision, GSI
Helmholtzzentrum f\"ur Schwerionenforschung GmbH, Planckstr.\ 1,
D-64291 Darmstadt, Germany
}%
\affiliation{Frankfurt Institute for Advanced Studies FIAS, Ruth-Moufang-Str. 1, D-60438 Frankfurt am Main, Germany
}%
\author{M. Elvers}
\affiliation{Institut f\"ur Kernphysik, Universit\"at zu K\"oln,
Z\"ulpicher Str. 77, D-50937 K\"oln, Germany
}%
\author{J. Endres}
\affiliation{Institut f\"ur Kernphysik, Universit\"at zu K\"oln,
Z\"ulpicher Str. 77, D-50937 K\"oln, Germany
}%
\author{M. Fritzsche}
\affiliation{%
Institut f\"ur Kernphysik, Technische Universit\"at Darmstadt, 
Schlossgartenstr. 9, D-64289 Darmstadt, Germany
}%
\author{B. L\"oher}%
\affiliation{ExtreMe Matter Institute EMMI and Research Devision, GSI
Helmholtzzentrum f\"ur Schwerionenforschung GmbH, Planckstr.\ 1,
D-64291 Darmstadt, Germany
}%
\affiliation{Frankfurt Institute for Advanced Studies FIAS, Ruth-Moufang-Str. 1, D-60438 Frankfurt am Main, Germany
}%
\author{N. Pietralla}
\affiliation{%
Institut f\"ur Kernphysik, Technische Universit\"at Darmstadt, 
Schlossgartenstr. 9, D-64289 Darmstadt, Germany
}%
\author{V. Yu. Ponomarev}
\affiliation{%
Institut f\"ur Kernphysik, Technische Universit\"at Darmstadt, 
Schlossgartenstr. 9, D-64289 Darmstadt, Germany
}%
\author{C. Romig}
\affiliation{%
Institut f\"ur Kernphysik, Technische Universit\"at Darmstadt, 
Schlossgartenstr. 9, D-64289 Darmstadt, Germany
}%
\author{L. Schnorrenberger}
\affiliation{%
Institut f\"ur Kernphysik, Technische Universit\"at Darmstadt, 
Schlossgartenstr. 9, D-64289 Darmstadt, Germany
}%
\author{K. Sonnabend}
\affiliation{%
Institut für Angewandte Physik, Goethe-Universit\"at Frankfurt, Max-von-Laue-Str. 1,
D-60438 Frankfurt am Main, Germany
}%
\author{A. Zilges}
\affiliation{Institut f\"ur Kernphysik, Universit\"at zu K\"oln,
Z\"ulpicher Str. 77, D-50937 K\"oln, Germany
}%

\date{\today}

\begin{abstract}

The low-lying electric dipole (E1) strength in the semi-magic nucleus
$^{136}$Xe has been measured which finalizes the systematic survey to
investigate the so-called pygmy dipole resonance (PDR) in all stable
even N=82 isotones with the method of nuclear resonance fluorescence
using real photons in the entrance channel. In all cases, a fragmented
resonance-like structure of E1 strength is observed in the energy
region 5 MeV to 8 MeV. An analysis of the fragmentation of the
strength reveals that the degree of fragmentation decreases towards
the proton-deficient isotones while the total integrated strength
increases indicating a dependence of the total strength on the
neutron-to-proton ratio. The experimental results are compared to
microscopic calculations within the quasi-particle phonon model
(QPM). The calculation includes complex configurations of up to three
phonons and is able to reproduce also the fragmentation of the E1
strength which allows to draw conclusions on the damping of the
PDR. Calculations and experimental data are in good agreement in the
degree of fragmentation and also in the integrated strength if the
sensitivity limit of the experiments is taken into account.

\end{abstract}


\maketitle

\section{Introduction}\label{sub::intro}

In strongly coupled many-body systems, collective excitations are a
common phenomenon expressing the strong interaction between the single
constituents.  In atomic nuclei, the classical example of collective
excitations are giant resonances. The theoretical description of the
damping of these collective modes within microscopic models is very
difficult because of their high excitation energies and different
mechanisms contributing to the damping width. The so-called Pygmy
Dipole Resonance (PDR), a concentration of electric dipole (E1) strength
below the well-known IsoVector Electric Giant Dipole Resonance (IVGDR)
\cite{hara01}, has attracted considerable interest during the last few
years. In contrast to the IVGDR, the PDR is an energetically low-lying
mode located often (partly) below the particle thresholds. Therefore,
coupling to complex configurations is the only mechanism for the
resonance damping. In addition, the density of complex configurations
in the energy region of the PDR is not too high allowing one to
account for nearly all of them in a microscopic model. Therefore, the
PDR is a challenge for theory in nuclear physics because one can
expect a good description of the fragmentation of a collective mode
without including any phenomenological parameters responsible for the
resonance width. Also from the experimental point of view the lower
energy of the PDR is advantageous since below the particle thresholds
the high resolution $\gamma$-ray spectroscopy in nuclear resonance
fluorescence (NRF) experiments using high-purity Germanium detectors
provides the necessary energy resolution in order to study the
fragmentation of the resonance. Thus, in the case of the PDR the
comparison of experimental data to the results of microscopic models
allows to determine whether the damping mechanism for a collective
mode in many-body systems is well understood.

In the last decade, the PDR has been established as a new type of E1
excitation both experimentally as well as in theory. Many modern
microscopic model calculations show an enhancement of the E1 strength
at low excitation energies on top of the tail of the IVGDR, see
%e.g. \cite{paar07, tson08, lanz09, litv09} 
e.g. \cite{paar07, endr10} and references therein. In most models the
PDR exhibits the signature of an oscillation of excess neutrons
against a proton-neutron core and, thus, the strength increases for
more neutron-rich nuclei.

Experimental evidence for a structure of strong low-lying E1 strength
in addition to the high energetic IVGDR has been found in many medium
heavy to heavy nuclei. After first indications found already three
decades ago \cite{bart73,metz78a,metz78b} and following results with
high-resolution and improved sensitivity
\cite{jung95,herz97,gova98,ryez02,zilg02} an experimental survey using
the method of NRF has been performed in different mass regions in
order to study the systematics of the PDR in stable nuclei
\cite{ende03,hart04,volz06,schw07,savr08b,ruse09,tonc10,maki10}. NRF
experiments are an ideal tool to investigate bound E1 strength
\cite{knei96}. Due to the low-momentum transfer of real photons in the
energy region of the PDR nearly exclusively $J=1$ states are populated
from the ground state in even-even nuclei. Thus NRF experiments
provide an excellent selectivity to the excitations of interest. Using
high-purity Germanium (HPGe) detectors for the $\gamma$-ray
spectroscopy of the emitted photons in the $(\gamma ,\gamma ')$
reaction an excellent energy resolution of less than 10 keV can be
achieved which allows the investigation of the fine structure and
fragmentation of the E1 strength as we present in this paper. In
addition, due to the well-known excitation mechanism the measured
observables can be directly linked to intrinsic properties of the
excited states in a model-independent way.

Currently NRF experiments using real photons are limited to stable
nuclei. The E1 strength in unstable nuclei can be studied using the
method of Coulomb excitations in inverse kinematics \cite{Auma05},
which allows to extend the systematics to very neutron-rich
systems. Results for neutron-rich Sn and Sb isotopes
\cite{adri05,klim07} and $^{68}$Ni \cite{wiel09} show an enhancement
of the strength located in the PDR region compared to the less
neutron-rich stable isotopes. This enhancement with the
neutron-to-proton ratio is an important evidence for the correlation
of the PDR strength to the neutron excess. Unfortunately the energy
resolution in these experiments does not allow the investigation of
the fine structure of the E1 strength and, thus, such investigations
will only be possible for stable isotopes in the near future.

In this paper, we report on the results of a NRF experiment on
$^{136}$Xe which finalizes a systematic investigation of the
properties of the PDR in the stable N=82 isotones below the neutron
separation energy. The results are compared in detail to calculations
within the quasi-particle phonon model (QPM) with a special focus on
the fragmentation of the strength. As mentioned in the first paragraph
the fragmentation itself provides an important observable to compare
to theory but it also has consequences on integral quantities as the
total integrated strength. First results of this comparison for
$^{136}$Xe have recently been reported in a letter \cite{savr08b}. In
this paper we extend the investigation of the fragmentation to all
stable even N=82 isotones and present detailed information on the
experiment on $^{136}$Xe. In the next section the experiment and the
data analysis are described, while Sec. \ref{sub::results} summarizes
the experimental results. Details on the QPM calculations are given in
Sec. \ref{sub::qpm}. In Section \ref{sub::compare}, a comparison of
the experimental results and the QPM calculations with respect to the
fragmentation and systematics for investigated N=82 isotones is
presented.

\section{Setup and data analysis}\label{sub::exp}

The experiments were performed at the Darmstadt High Intensity Photon
Setup (DHIPS) using bremsstrahlung as photon source
\cite{sonn11}. Bremsstrahlung is produced by completely stopping the
electron beam of the injector module of the superconducting electron
linear accelerator S-DALINAC \cite{brun99} in a copper radiator where
electron beam intensities of about 40 $\mu$A and up to 10 MeV are
available \cite{baye07}. The photon beam is collimated to a size of
about 2 cm diameter at the target position. Three large-volume HPGe
detectors, each equipped with a BGO anti-Compton shield and surrounded
with heavy passive shielding, have been used for the spectroscopy of
the high-energy photons emitted in the $(\gamma ,\gamma ')$
reaction. More details on the setup can be found in \cite{sonn11}.

In the present study, two measurements have been performed. The first
one using bremsstrahlung with an endpoint energy (defined by the
electron energy) of 9.2 MeV to ensure a sufficient photon flux up to
the neutron separation energy of $^{136}$Xe ($S_{n} = 8.06$ MeV) and
another one at 8.0 MeV just below $S_{n}$ to identify possible
contributions from $(n,\gamma )$ reactions of neutrons produced by the
photo-dissociation of $^{136}$Xe in the 9.2 MeV measurement.

The NRF target consisted of four high-pressure gas containers made of
Titanium and filled with Xenon enriched to 99.9 \% in the isotope of
interest. Similar targets have been used in \cite{garr06} for previous
NRF experiments on Xenon isotopes up to 4 MeV. In total, 2.925 g of
$^{136}$Xe was enclosed in the four containers. Further details on the
high-pressure gas targets can be found in
\cite{reif02,garr06,rupp09}. For the calibration of energy, photon
flux and efficiency the $^{136}$Xe was sandwiched between two $^{11}$B
targets with a total mass of 0.77 g. The combined target was
smaller than the photon beam diameter and, thus, homogeneously
irradiated.

\begin{figure}
\includegraphics[width=\columnwidth]{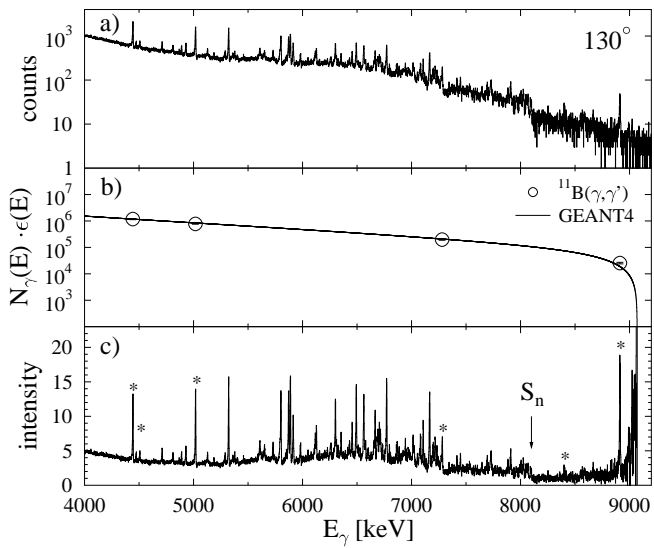}
\caption{\label{fig::spec} a) Measured NRF spectrum for one detector
at 130$^{\circ}$ with respect to the incoming photon beam at an
endpoint energy of 9.2 MeV. b) Determined product of photon flux
$N_{\gamma}(E)$ and photopeak efficiency $\epsilon(E)$. The simulated
shape has been normalized using the well-known transition in the
$^{11}$B$(\gamma ,\gamma ')$ reaction. c) Measured spectrum normalized
to $N_{\gamma}(E) \cdot \epsilon (E)$. Peak areas in this intensity
spectrum are directly proportional to integrated cross sections. Peaks
corresponding to transitions in the normalization standard $^{11}$B
are marked with an asterisk (*).}
\end{figure}

An example of a measured $\gamma$-ray energy spectrum is shown in the
upper part of Fig. \ref{fig::spec} for an endpoint energy of 9.2
MeV. The spectrum has been obtained with one detector at an angle of
130$^{\circ}$ with respect to the incoming beam direction. Besides
peaks stemming from the $^{11}$B$(\gamma ,\gamma ')$ reaction and a
few background lines all peaks correspond to transitions in
$^{136}$Xe.

In NRF experiments, the measured area $A_{i}$ of a peak corresponding
to the elastic scattering channel of an excited state (i.e. decay back to
the ground state) after photo-excitation from the ground state is
related to the energy-integrated cross section $I_{i,0}$ by
\begin{align}\label{eq::peak}
A_{i} \propto I_{i,0} \cdot W_{i}(\theta ,\phi ) \cdot
N_{\gamma}(E_{i}) \cdot \epsilon (E_{\gamma})
\end{align}
with $W_{i}(\theta ,\phi )$ denoting the angular distribution of the
emitted photons, $N_{\gamma}(E_{i})$ the incident photon intensity and
$\epsilon (E_{\gamma})$ the absolute photo-peak efficiency of the
corresponding detector, respectively. The angular distribution function
$W_{i}(\theta ,\phi )$ depends exclusively on the multipole
decomposition of the $\gamma$-ray transitions between the involved
states and can thus be used to determine the multipolarities of the
transitions and the spin quantum numbers of the excited states.

For an absolute determination of the integrated cross section
$I_{i,0}$, the product of photon intensity $N_{\gamma} (E_{i})$ and
photo-peak efficiency $\epsilon(E_{\gamma})$ has to be calibrated. For
transitions back to the ground state (i.e. $E_{i} = E_{\gamma}$), this
product can be directly determined for the energies of the excited
states of $^{11}$B since their NRF cross sections are known with good
accuracy \cite{Ajze90}. In order to interpolate between these
energies, simulations of the bremsstrahlung spectrum and of the
photopeak efficiency are performed using GEANT4
\cite{geant03,geant06}. It has been shown, that GEANT4 reproduces the
energy dependence of our bremsstrahlung beam for the present energy
region with good accuracy \cite{sonn11}. Finally, the product of the
simulated photon spectrum and efficiency is fitted to the experimental
values at the $^{11}$B energies in order to yield an absolute
calibration for the entire energy region. The middle part of
Fig. \ref{fig::spec} shows the result of this procedure.

Since photon intensity and detection efficiency are decreasing
strongly for higher photon energies the decrease in measured spectrum
towards the neutron separation energy might be due to this reason. In
order to correct for this effect, the measured spectrum is divided by
the determined energy dependence of $N_{\gamma}(E_{i}) \cdot \epsilon
(E_{i})$. The resulting intensity spectrum is shown in the bottom part
of Fig. \ref{fig::spec}. Clearly, the spectrum is dominated by the
strong transitions in the energy region of about 5.5 to 7.5 MeV.

\begin{figure}
\includegraphics[width=\columnwidth]{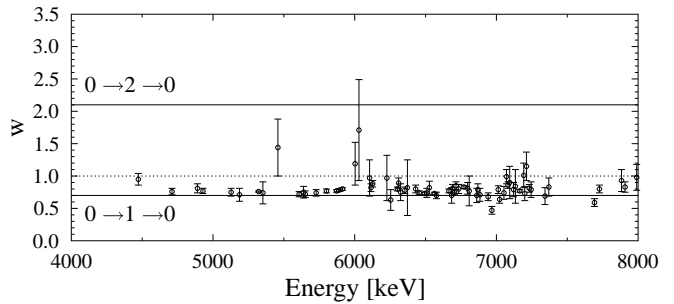}
\caption{\label{fig::ang} Measured angular distributions for all
observed transitions in $^{136}$Xe in the energy region 4 MeV to 8
MeV. Except for few weak states $J=1$ can be assigned in all cases.}
\end{figure}

As mentioned above the spin of the excited states can be determined by
measuring the angular distribution of the emitted photons. Figure
\ref{fig::ang} shows the ratios
\begin{align}\label{eq::ang}
W = \frac{A_{i}(90^{\circ})}{A_{i}(130^{\circ})} \cdot \frac{\epsilon_{130^{\circ}}(E_i)}{\epsilon_{90^{\circ}}(E_i)}
\end{align} 
of the measured peak areas $A_{i}(90^{\circ})$ and
$A_{i}(130^{\circ})$ above 3.5 MeV (corrected for the individual
efficiencies) of the detectors at 90$^{\circ}$ and 130$^{\circ}$,
respectively. The expectation values for transitions back to the
ground state, i.e. the spin sequences $0_{1}^{+} \rightarrow 1
\rightarrow 0_{1}^{+}$ and $0_{1}^{+} \rightarrow 2 \rightarrow
0_{1}^{+}$, are indicated by horizontal lines. Except for a few
very weak transitions dipole character and thus $J=1$ for the excited
states can be assigned unambiguously.

No information on the parities of the excited states could be obtained
from the present data. However, investigations in the neighboring
nuclei $^{138}$Ba \cite{piet02,tonc10} and $^{140}$Ce \cite{bues08}
have shown that the by far dominant part of the dipole strength has
E1 character. Thus the assumption of E1 character in the case of
$^{136}$Xe seems to be very likely and only a small M1 contribution
can be expected.

An advantage of the NRF method is the fact that transition widths
can be extracted from measured cross sections in a model-independent way
\cite{knei96}. The connection to the integrated cross section $I_{i,0}$ is given by 
\begin{align}\label{eq::widths}
I_{i,0} = \pi^{2} \left( \frac{\hbar c}{E_{i}} \right)^{2} \cdot g
\cdot \frac{\Gamma_{0}^{2}}{\Gamma}
\end{align}
with $\Gamma_{0}$ and $\Gamma$ being the ground state and total decay
width, respectively. The spin factor \mbox{$g = (2 J_{i} + 1)/(2 J_{0}
+1)$} is given by the spin of ground-state and the excited state the
latter being determined by the measured value of $W$ (compare
Eq. \ref{eq::ang}). Equation (\ref{eq::widths}) enables to determine
the absolute ground-state decay-width $\Gamma_{0}$ from the measured
integrated cross section if the branching ratio $\Gamma_{0}/\Gamma$ is
known. In the present study, no transitions corresponding to inelastic
decay channels of the identified $J=1$ states could be observed. Thus,
$\Gamma_{0}/\Gamma = 1$ is assumed in all cases. Since many weak (and
thus unobserved) branchings might add up one should keep in mind that
$\Gamma_{0}/\Gamma$ might be somewhat smaller and the extracted values
for $\Gamma_{0}$ represent a lower limit. However, in \cite{tonc10} it
has been shown for $^{138}$Ba that the averaged contribution of
inelastic decays to the photo absorption cross section is small for
excitation energies below 7 MeV and becomes significant only at higher
energies. Therefore, $\Gamma_{0}/\Gamma$ seems to be close to unity
and the above made assumption is reasonable.

Using the determined ground state decay widths the $B(E1)\uparrow$
transition strength can be calculated for each individual state:
\begin{align}\label{eq::be1}
\frac{B(E1) \uparrow}{\rm{e^{2}fm^{2}}} = 9.554 \cdot 10^{-4} \cdot g
\cdot \frac{\Gamma_{0}}{\rm{meV}} \cdot \left( \frac{\rm{MeV}}{E_{i}}
\right)^{3}.
\end{align}
In the following chapter the results of the experiment on $^{136}$Xe
are presented and compared to the other N=82 isotones.

\section{Experimental results}\label{sub::results}

Using the method described in the previous section the integrated
cross section as well as the spin of 76 newly observed states up to
the neutron separation energy are determined from the measured NRF
spectra for $^{136}$Xe. Only in a few cases the angular distribution is
also consistent with a $J=2$ assignment in all other cases $J=1$ can
be clearly assigned (compare Fig. \ref{fig::ang}). Assuming
$\Gamma_{0}/\Gamma = 1$ (see discussion in the previous section) we
obtain the corresponding ground-state decay-widths $\Gamma_{0}$ that are
propotional to the excitation strengths $B(E1)$$\uparrow$ in case of
negative parity for the excited $J=1$ states. The results are
summarized in Tab. \ref{tab::136xe}. No systematic deviations have
been observed for the obtained cross sections between the measurements
at the two different endpoint energies and, thus, the results for the
single experiments have been combined.

\begingroup 
\squeezetable
\begin{table}
\caption{\label{tab::136xe} Results of the experiment on
$^{136}$Xe. Given are deduced energy, angular distribution, spin,
partial decay width and $B(E1)\uparrow$ strength of the observed
states. For the latter $\pi = -1$ and $\Gamma_{0}/\Gamma = 1$ was
used, since no branching transitions were observed.}
\begin{tabular}{p{1.5cm}cccc}
\\[-0.1cm]
\hline
\hline
\\[-0.2cm]
$E_{x}$ & W & $J^{\pi}$ & $\Gamma_{0}^{2} / \Gamma$ & $B(E1) \uparrow$            \\
(keV)   &   & \quad $(\hbar)$ \, \,& (eV)                     & ($10^{-3}$ e$^{2}$fm$^{2}$) \\[0.1cm]
\hline
\\[-0.2cm]
    3626(1)\footnote[1]{also observed in \cite{garr06} with B(E1) = $1.52(12) \cdot 10^{-3}$ e$^{2}$fm$^{2}$} & 1.20(45) & 1,2 &  0.027(6) &         1.7(3)\footnotemark[1] \\ 
    3738(1) & 0.93(13) & 1 &        0.057(5) &         3.1(3) \\ 
    4473(1) & 0.95(9) & 1 &         0.120(8) &         3.9(3) \\ 
    4711(1) & 0.76(5) & 1 &         0.217(13) &         6.0(4) \\ 
    4890(1) & 0.81(7) & 1 &         0.175(11) &         4.3(3) \\ 
    4929(1) & 0.77(4) & 1 &         0.341(19) &         8.1(5) \\ 
    5128(1) & 0.75(6) & 1 &         0.199(13) &         4.2(3) \\ 
    5187(1) & 0.71(10) & 1 &        0.088(8) &         1.8(2) \\ 
    5322(1) & 0.76(1) & 1 &         1.66(9) &        31.6(16) \\ 
    5352(1) & 0.74(17) & 1 &        0.072(9) &         1.3(2) \\ 
    5458(1) & 1.44(44) & 1,2 &      0.122(17) &         2.2(3) \\ 
    5608(1) & 0.72(4) & 1 &         0.491(29) &         8.0(5) \\ 
    5639(1) & 0.75(9) & 1 &         0.210(17) &         3.4(3) \\ 
    5651(1) & 0.72(5) & 1 &         0.441(26) &         7.0(4) \\ 
    5728(1) & 0.74(5) & 1 &         0.382(24) &         5.8(4) \\ 
    5801(1) & 0.77(3) & 1 &         2.02(11) &        29.7(16) \\ 
    5872(1) & 0.77(2) & 1 &         2.05(11) &        29.1(15) \\ 
    5888(1) & 0.78(2) & 1 &         2.59(13) &        36.3(19) \\ 
    5914(1) & 0.80(2) & 1 &         1.36(7) &        18.9(10) \\ 
    6003(1) & 1.19(33) & 1,2 &      0.126(18) &         1.7(2) \\ 
    6030(1) & 1.71(78) & 1,2 &      0.088(16) &         1.1(2) \\ 
    6105(1) & 0.97(28) & 1 &        0.159(26) &         2.0(3) \\ 
    6115(1) & 0.82(6) & 1 &         0.661(48) &         8.3(6) \\ 
    6127(1) & 0.88(5) & 1 &         1.10(7) &        13.7(9) \\ 
    6227(1) & 0.97(35) & 1 &        0.304(50) &         3.6(6) \\ 
    6254(1) & 0.63(16) & 1 &        0.236(32) &         2.8(4) \\ 
    6301(1) & 0.80(3) & 1 &         1.99(11) &        22.8(12) \\ 
    6310(1) & 0.89(8) & 1 &         0.629(42) &         7.2(5) \\ 
    6324(1) & 0.75(13) & 1 &        0.214(22) &         2.4(2) \\ 
    6354(1) & 0.78(5) & 1 &         0.796(47) &         8.9(5) \\ 
    6372(1) & 0.82(43) & 1 &        0.108(30) &         1.2(3) \\ 
    6430(1) & 0.80(6) & 1 &         0.643(40) &         6.9(4) \\ 
    6455(1) & 0.74(3) & 1 &         1.38(8) &        14.7(8) \\ 
    6493(1) & 0.73(2) & 1 &         2.62(14) &        27.5(14) \\ 
    6509(1) & 0.74(7) & 1 &         0.489(34) &         5.1(4) \\ 
    6527(1) & 0.82(10) & 1 &        0.410(32) &         4.2(3) \\ 
    6562(1) & 0.73(2) & 1 &         2.43(13) &        24.6(13) \\ 
    6577(1) & 0.70(5) & 1 &         0.703(43) &         7.1(4) \\ 
    6665(1) & 0.77(3) & 1 &         1.81(10) &        17.5(10) \\ 
    6684(1) & 0.70(12) & 1 &        0.482(47) &         4.6(4) \\ 
    6691(1) & 0.81(6) & 1 &         1.19(8) &        11.4(7) \\ 
    6704(1) & 0.75(4) & 1 &         1.32(8) &        12.5(7) \\ 
    6715(1) & 0.85(6) & 1 &         1.17(7) &        11.0(7) \\ 
    6734(1) & 0.80(7) & 1 &         0.673(44) &         6.3(4) \\ 
    6771(1) & 0.83(2) & 1 &         3.46(18) &        31.9(17) \\ 
    6797(1) & 0.81(9) & 1 &         0.542(40) &         4.9(4) \\ 
    6808(1) & 0.77(23) & 1 &        0.190(28) &         1.7(3) \\ 
    6861(1) & 0.70(11) & 1 &        0.407(38) &         3.6(3) \\ 
    6869(1) & 0.79(9) & 1 &         0.621(48) &         5.5(4) \\ 
    6884(1) & 0.71(10) & 1 &        0.342(29) &         3.0(3) \\ 
    6942(1) & 0.68(6) & 1 &         0.668(45) &         5.7(4) \\ 
    6968(1) & 0.47(6) & 1 &         0.338(30) &         2.9(3) \\ 
    7013(1) & 0.79(06) & 1 &        1.01(6) &         8.4(5) \\ 
    7023(1) & 0.64(06) & 1 &        0.683(47) &         5.6(4) \\ 
    7053(1) & 0.74(10) & 1 &        0.398(33) &         3.3(3) \\ 
    7071(1) & 0.99(11) & 1 &        0.670(49) &         5.4(4) \\ 
    7082(1) & 0.86(05) & 1 &        1.44(8) &        11.6(7) \\ 
    7094(1) & 0.90(25) & 1 &        0.283(44) &         2.3(4) \\ 
    7121(1) & 0.79(11) & 1 &        0.394(34) &         3.1(3) \\ 
    7134(1) & 0.84(26) & 1 &        0.247(42) &         1.9(3) \\ 
    7165(1) & 0.77(2) & 1 &         3.43(18) &        26.7(14) \\ 
    7193(1) & 1.01(19) & 1 &        0.74(8) &         5.7(6) \\ 
    7200(1) & 0.73(11) & 1 &        0.84(8) &         6.5(6) \\ 
    7212(1) & 1.15(22) & 1 &        0.91(9) &         6.9(7) \\ 
    7232(1) & 0.81(6) & 1 &         1.00(7) &         7.6(5) \\ 
    7245(1) & 0.79(12) & 1 &        0.361(38) &         2.7(3) \\ 
    7343(1) & 0.69(13) & 1 &        0.292(31) &         2.1(2) \\ 
    7370(1) & 0.83(14) & 1 &        0.357(35) &         2.6(3) \\ 
    7692(1) & 0.59(6) & 1 &         0.78(6) &         4.9(4) \\ 
    7727(1) & 0.80(6) & 1 &         1.69(11) &        10.5(7) \\ 
    7883(1) & 0.93(17) & 1 &        0.80(9) &         4.7(5) \\ 
    7908(1) & 0.83(8) & 1 &         1.65(14) &         9.6(8) \\ 
    7990(1) & 0.98(20) & 1 &        0.75(9) &         4.2(5) \\ 
    8024(1) & 0.76(8) & 1 &         1.40(12) &         7.8(7) \\ 
    8051(1) & 0.93(13) & 1 &        1.11(11) &         6.1(6) \\ 
    8066(1) & 0.75(12) & 1 &        0.86(9) &         4.7(5) \\ 
    8093(1) & 0.77(10) & 1 &        1.02(10) &         5.5(5) \\ 
\hline
\hline
\end{tabular}
\end{table}
\endgroup

In a previous NRF experiment \cite{garr06} on $^{136}$Xe with a
bremsstrahlung endpoint energy of 4 MeV only two states have been
observed at 2869 keV and 3626 keV with B(E1) values of $0.169(23)
\cdot 10^{-3}$ e$^{2}$fm$^{2}$ and $1.52(12) \cdot
10^{-3}$ e$^{2}$fm$^{2}$, respectively. While the lower lying state is
below the sensitivity limit of the present experiment the higher lying
one is observed and a B(E1) strength of $1.7(3) \cdot
10^{-3}$ e$^{2}$fm$^{2}$ has been determined. The good agreement of the
two values excludes a strong feeding component in the present study
even though the endpoint energy of the bremsstrahlung is above the
neutron threshold.

Unlike in the other N=82 isotones no clear and isolated candidate for
the $[2^{+} \otimes 3^{-}]_{1^{-}}$ two-phonon state is observed in
$^{136}$Xe. Following the systematics of the two-phonon states in the
other N=82 isotones \cite{andr01} such a state would be expected
around 4.2 MeV with a strength in the order of B(E1) $\approx 10 \cdot
10^{-3}$ e$^{2}$fm$^{2}$. Around 4.5 MeV and slightly higher energies
a few states are observed in $^{136}$Xe that however have smaller
individual B(E1) strengths than expected for the two-phonon
state. Thus, the $[2^{+} \otimes 3^{-}]_{1^{-}}$ strength is either
fragmented into several states that share the two-phonon E1 strength
or its strength is further reduced.

\begin{figure}
\includegraphics[width=\columnwidth]{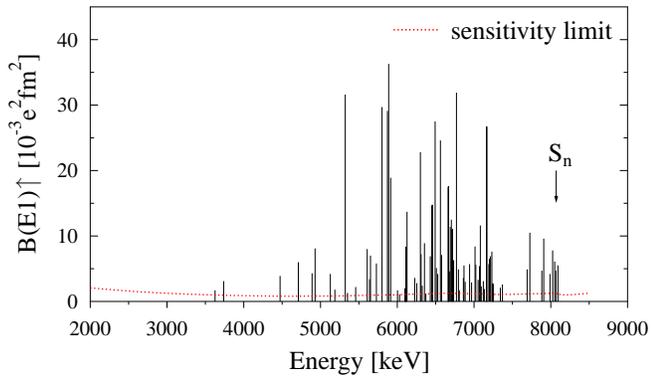}
\caption{\label{fig::xenonBE1} (Color online) Measured E1 strength
distribution in $^{136}$Xe up to the neutron separation energy. The
dotted line indicates the sensitivity limit of the experiment. }
\end{figure}

The measured B(E1) strength distribution for $^{136}$Xe up to $S_{n}$
is shown in the upper part of Fig. \ref{fig::xenonBE1}. The
sensitivity limit of the experiment is indicated by the dotted line
and has been calculated based on the continuous background present in
the spectra and following the formalism given in \cite{endr09a} using
a confidence limit of $3\sigma$. Due to this finite sensitivity
excitations with smaller $B(E1) \uparrow$ values thus will remain
unobserved.

\begin{figure*}
\includegraphics[width=\textwidth]{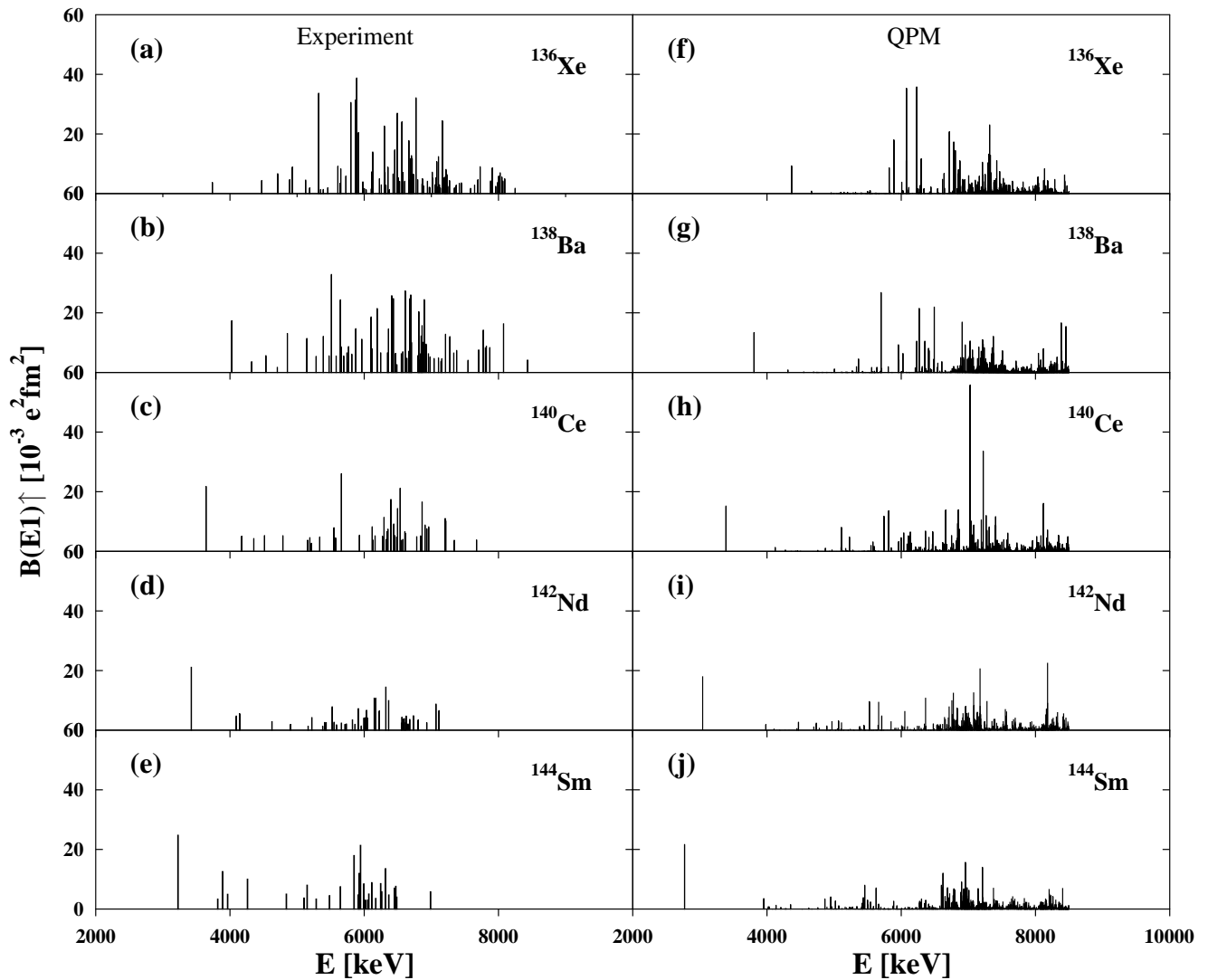}
\caption{\label{fig::n82BE1} Measured (left side) and calculated
(right side) B(E1) strength distributions in the stable even N=82
isotones. The experimental sensitivity limits of all experiments are
similar to the one shown for $^{136}$Xe in
Fig. \ref{fig::xenonBE1}. Data taken partly from \cite{volz06}. }
\end{figure*}

Similar as in the other stable N=82 isotones also in $^{136}$Xe a
concentration of strong E1 excitations in the energy region of 5 MeV to 7 MeV
is observed. The left part of Fig. \ref{fig::n82BE1} shows the
measured B(E1) strength distributions for all N=82 isotones. The
distribution of $^{136}$Xe fits into the systematics with an
increasing strength and also increasing excitation energy compared to
the heavier isotones. A more detailed comparison of the B(E1) strength
distribution in terms of total strength and also fragmentation is
presented in Sec. \ref{sub::compare}.

The facts that the experimental sensitivity is always limited and, in
addition, branching transitions are difficult to observe consequently
lead to an incomplete determination of the total E1 strength using the
presented experimental technique. Recently, attempts have been made to
extract also the strength in the continuum part of the experimental
spectra and in parallel to account for the inelastic decay branchings
in order to estimate the total photo absorption cross section
\cite{schw08,ruse09,maki10}. However, in this approach one loses the
model independence of the NRF method and also highly relies on very
accurate simulations of background and detector response
functions. Therefore, we do not follow this approach here especially
since the contribution of unresolved strength in the case of the
semi-magic $^{136}$Xe isotope can be expected to be lower and the
inelastic contribution has been shown to be rather small in the
neighboring nucleus $^{138}$Ba \cite{tonc10} in the energy region of
interest. Even though the total strength extracted will then represent
a lower limit, the result remains model independent which is one of
the great advantages of real photon scattering experiments and still
allow for a solid and unambiguous comparison to theory as it is done
below.

In a comparison to model calculations this limited knowledge about the
total excitation strength needs to be accounted for. This can be done
by comparing only the part of the strength produced in a model
calculation which is above the experimental sensitivity and, thus, can
be expected to be observed in the experiment. Of course, to allow such
a kind of comparison the fragmentation of the E1 strength has to be
reproduced correctly in the model in order to account for the
experimental sensitivity. The QPM calculations presented in the next
section fulfill this requirement as we will show in Sec.
\ref{sub::compare}.

\section{QPM calculations}\label{sub::qpm}

Excited states of even-even nuclei are treated in the quasi-particle
phonon model (QPM) \cite{solo92} in terms of phonons with spin and
parity $\lambda^{\pi}$; the ground state is considered as a phonon
vacuum.  The phonons are made up of quasi-particle pairs. Their spectra
and internal fermion structure is obtained by solving the QRPA
equations for each multipolarity. The QRPA involves $0p4h$, $2p2h$,
and $4p0h$ terms of the residual two-body interaction. This
interaction in the QPM has a simple separable form. The remaining
$1p3h$ and $3p1h$ terms of the residual interaction are responsible
for the mixing between one- and two-, two- and three-, etc. phonon
configurations. Accordingly, the wave function of excited states in
the QPM is written as a composition of one-, two-, etc phonon
configurations. The energies of excited states and components of their
wave functions are found from a diagonalization of the model Hamiltonian
on the set of these wave functions.

Although the QPM wave functions have a complex form their one-phonon
components play a decisive role in the excitation process of these
states from the ground state by an external field (e.g.,
electromagnetic) described by a one-body operator. Two-phonon
components are also excited from the QRPA ground state. An example is
the first $1^-$ state in spherical nuclei which has almost pure
$[2^+_1 \otimes 3^-_1]_{1^-}$ nature. But in general, their transition
matrix elements are much smaller compared to the ones of one-phonon
components. Thus, complex (two-, three-phonon) configurations
participate in the creation of the fragmentation pattern of the
excitation strength but add very little to the total
strength. 

\begin{figure}[h] \includegraphics[width=\columnwidth]{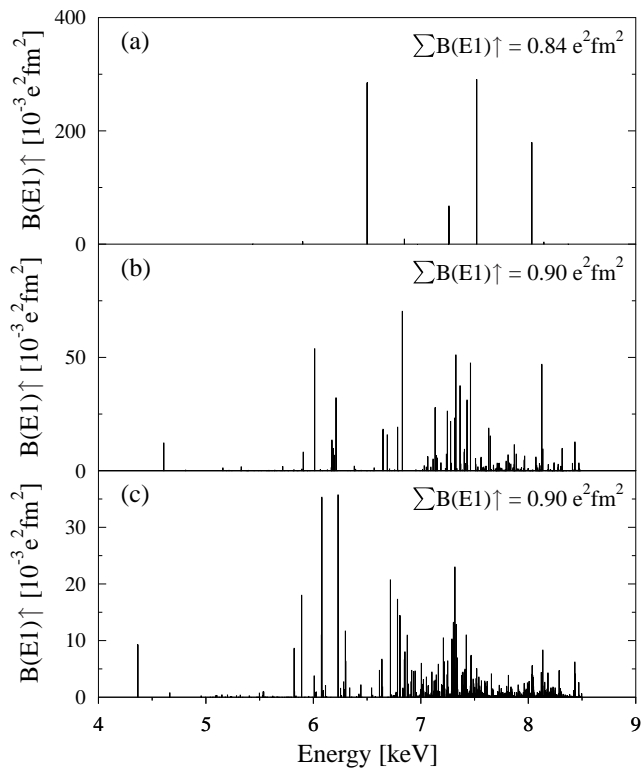}
\caption{\label{fig::frag} The QPM calculations of the B(E1) strength
distribution in $^{136}$Xe performed a) in one-phonon approximation
and with the wave function containing b) one- and two- and c) one-,
two-, and three-phonon configurations. The summed strength is hardly
influenced while the fragmentation increases rapidly.}
\end{figure}

The formation of the fragmentation pattern is demonstrated in
Fig.~\ref{fig::frag} in which the distribution of the E1 strength of
the PDR in $^{136}$Xe is presented. Figure~\ref{fig::frag}a presents
the results obtained in the one-phonon approximation. One notices that
the E1 strength in this energy region originates from four one-phonon
states. Calculations performed with the wave function containing one-
and two-phonon configurations are shown in Fig.~\ref{fig::frag}b. The
B(E1) value for each individual $1^-$ state drops dramatically as
compared to the results in Fig.~\ref{fig::frag}a. This is due to the
fact that the number of two-phonon configurations in this energy
interval is much larger. Interaction between one- and two-phonon
configurations leads to their mixing and the contribution of the
one-phonon configurations (which carry E1 strength) to the wave
function norm does not exceed a few percents for each state. Note
also the appearance of the two-phonon $1^-$ state discussed above at
around 4.5~MeV in this step. The fragmentation progresses further when three-phonon
configurations are added (Fig.~\ref{fig::frag}c). Many states with
rather small B(E1) values appear especially at higher energies with
rapid increase of the density of three-phonon configurations.

The QPM calculations of the fine structure of the PDR in the
N=82 isotones in this paper extend previous calculations in $^{138}$Ba
\cite{herz99} and $^{140}$Ce \cite{herz97} which together with
Ref.~\cite{gova98} were the first theoretical studies on the topic.
Compared to the old calculations we have enlarged the basis of complex
configurations: two- and three-phonon configurations have been built up from
phonons with the multipolarity from $1^{\pm}$ up to $9^{\pm}$ with the
excitation energy below 8.5~MeV. Since the density of four-phonon
configurations (not included in the wave function) is still very low at
these energies, our basis is almost complete. The number of complex
configurations fluctuates slightly from nucleus to nucleus. On average, our
basis contains about 350 two-phonon and 900 three-phonon configurations. 
To account for admixture of the GDR in the low-energy region, all 1$^-$
one-phonon configurations below 20~MeV have been included in the wave
function of excited states.

The calculations in all N=82 isotones have been performed with the
same mean field which has been described by the Woods-Saxon potential
with parameters taken from a global parametrization~\cite{pono79} and
the same monopole pairing strength. Single-particle energies of the
mean field near the Fermi surface have been corrected to reproduce the
experimentally known single-particle levels in neighboring odd-mass
nuclei in the calculations with the wave function containing
``[quasiparticle~$\otimes$~N phonon]'' ($N=0,1,2,3$) components.  The
strength parameters of the residual interaction have been adjusted in
each particular nucleus in accordance with a standard QPM procedure
(see, e.g., \cite{bert99}).

\begin{figure}
\includegraphics[width=\columnwidth]{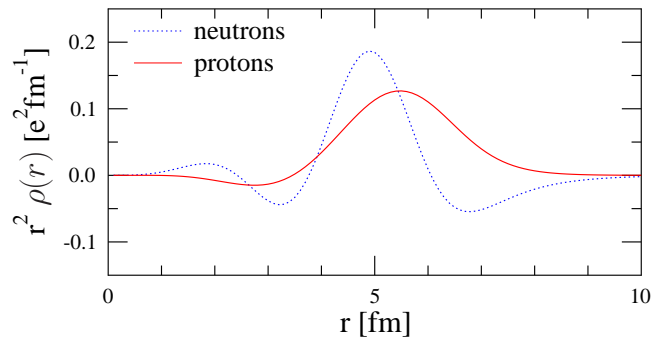}
\caption{\label{fig::density} (Color online) Calculated transition
densities for protons (red) and neutrons (blue) in the PDR region for
$^{136}$Xe.}
\end{figure}

Figure \ref{fig::density} shows calculated transition densities
separately for protons an neutron for the PDR region in
$^{136}$Xe. Very similar transition densities are obtained for the
other N=82 isotones and are thus not shown here.

\section{Comparison}\label{sub::compare}

A comparison between the experimentally determined B(E1) strength
distributions and the results of the QPM calculations for all stable
even N=82 isotones is shown in Fig. \ref{fig::n82BE1}. Besides a shift
of about 500 keV there is in general a good agreement between
experiment and QPM in the absolute scale of the strongest excitations
and the shape of the distributions. In all cases a resonance-like
structure is observed in experiment as well as in the calculation
which increases in energy and strength from $^{144}$Sm to
$^{136}$Xe.

\begin{figure}
\includegraphics[width=\columnwidth]{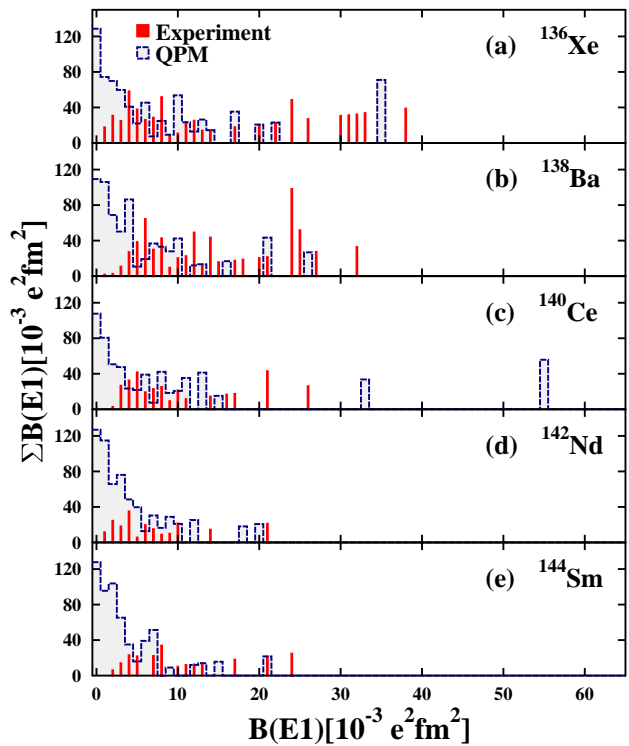}
\caption{\label{fig::binBE1} (Color online) Analysis of the
fragmentation in the experimentally observed distribution and within
the QPM model for the N=82 isotones. For details see text.}
\end{figure}

In order to compare the fragmentation of the E1 strength produced
within the QPM with the experimental result the B(E1) strength
distribution is shown in Fig. \ref{fig::binBE1} as a function of the
strength of the individual states. To produce Fig. \ref{fig::binBE1},
for each nucleus the states have been grouped according to their B(E1)
strengths in bins of $1 \cdot 10^{-3}$ e$^{2}$fm$^{2}$ and the summed
$\sum$B(E1) strength is calculated for each bin. Figure
\ref{fig::binBE1} thus shows how the strength is distributed
(fragmented) over the individual states. In this way, the limited
experimental sensitivity is expressed by the fact that no or only
little strength is observed for the group of states with very small
($\le 2 \cdot 10^{-3}$ e$^{2}$fm$^{2}$) strength. For these bins, the
experiment clearly misses strength. However, for stronger states the
experimental results can be expected to be complete and thus, this
part needs to be compared to the QPM calculation. For the region above
about $3 \cdot 10^{-3}$ e$^{2}$fm$^{2}$ experiment and calculation
show indeed very similar distributions. The QPM therefore nicely
reproduces the experimental distribution, i.e. the fragmentation is
described correctly within the model. This confirms that the damping
mechanism of the PDR is the coupling of a collective mode with complex
configurations. For the weaker excitations the QPM predicts much more
strength than is observed experimentally which clearly is a sign of
the limited experimental sensitivity. However, since the QPM
reproduces the experimental results for stronger states, the amount of
strength located in states with small B(E1) values in the QPM can be
expected to be a good estimate for the missing experimental strength.

Both, experiment and QPM, show a decreasing amount of excitations with
large B(E1) strength with increasing proton number. While in
$^{136}$Xe several states with B(E1) values well above $20 \cdot
10^{-3}$ e$^{2}$fm$^{2}$ exist, there is only one in $^{142}$Nd or
$^{144}$Sm. In general the strength is distributed among more and more
states with small E1 excitations strengths for increasing proton
number, i.e. the fragmentation increases. This is related to the fact
that the energy of the $3^-_1$ state systematically decreases from
$^{136}$Xe to $^{144}$Sm while the $2^+_1$ state has almost the same
energy in all isotones. In other words, collectivity of the $3^-_1$
state increases from light to heavy isotones making coupling matrix
elements between one-phonon $1^-$ configurations and many complex
configurations stronger.

\begin{figure}
\includegraphics[width=\columnwidth]{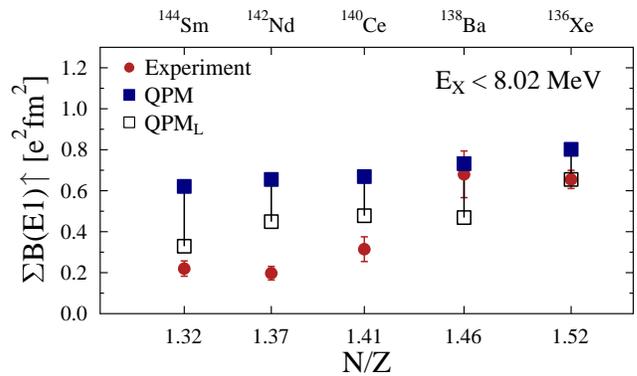} \caption{\label{fig::sumBE1}
(Color online) Integrated E1 strength for all stable even N=82 isotones up to an
excitation energy of 8.02 MeV ($S_{n}$ of $^{136}$Xe) as a function of
neutron-to-proton ratio N/Z. For the QPM calculations two values are
given, one including all states (QPM) and one including states above
the corresponding experimental sensitivity limits only (QPM$_{L}$).  }
\end{figure}

The integrated E1 strength up to 8.06 MeV (the neutron separation
energy of $^{136}$Xe) for the five investigated N=82 isotones is shown
in Fig. \ref{fig::sumBE1} as a function of the neutron-to-proton ratio
N/Z and summarized in Tab. \ref{tab::n82}. The experimental values
show a rather strong dependence on N/Z with values of 0.220(37)
e$^{2}$fm$^{2}$ for $^{144}$Sm and 0.662(45) e$^{2}$fm$^{2}$ for
$^{136}$Xe, respectively. This enhancement for larger N/Z is only
partly observed in the results of the QPM calculation which also shows
an increasing total strength but the dependency is much weaker. Also,
for the more proton-rich nuclei much more strength is predicted by the
QPM than observed experimentally.

Since the fragmentation of the strength is well reproduced by the QPM
it is now possible to study the effect of the limited experimental
sensitivity on this integral quantity.  For the values given by the
open squares (QPM$_{L}$) only the states above the experimental
sensitivity limit of the corresponding experiments have been included
in the total sum. The agreement between experiment and calculation is
clearly improved by accounting for the experimental sensitivity. Since
the fragmentation increases for higher proton numbers (and thus
smaller N/Z) the influence of the sensitivity limit becomes also more
important in these nuclei. The experimentally observed strong
enhancement of the total B(E1) strength for larger N/Z ratio, thus,
seems to be partly due to the higher fragmentation of the PDR in the
more proton-rich nuclei and the consequently larger amount of
experimentally unobserved strength. However, only part of the
enhancement is due to this effect and also the full QPM calculation
shows an enhancement of the total strength with increasing
neutron-to-proton ratio. This comparison implies that the
consideration of the experimental sensitivity limits plays an
important role in the systematics of integral quantities.

\begin{table*}
\caption{\label{tab::n82} Summary of results on the systematics in the stable even N=82 isotones. Energy values of the $2_{1}^{+}$ and $3_{1}^{-}$ states are taken from \cite{ensdf}. }
\begin{tabular}{p{1.5cm}c|ccc|ccc}
\\[-0.1cm]
\hline
\hline
\\[-0.2cm]
Isotope & N/Z & \quad E($2_{1}^{+}$) & \quad E($3_{1}^{-}$) & \quad E($1_{1}^{-}$) & \multicolumn{3}{c}{$\sum B(E1) \uparrow$ \footnote[1]{up to an excitation energy of E$ = 8.06$ MeV (S$_{n}$ of $^{136}$Xe).}}\\[0.1cm]
& &  \quad (MeV) \,& \quad (MeV) \,& \quad (MeV) \,& \multicolumn{3}{c}{(e$^{2}$fm$^{2}$)}\\[0.1cm]
\hline
\\[-0.2cm]        
& &  & &  & \, Exp.\, & \,\, QPM \,\,& \,\, QPM$_{L}$ \,\,\\[0.1cm]
\hline
\\[-0.2cm]      
$^{136}$Xe & 1.518  & \quad 1.313 \quad & \quad 3.275 \quad & \quad (3.738)\footnote[2]{Unknown parity.} \quad &  \quad 0.655(44) \quad & 0.802 & 0.655\\
$^{138}$Ba & 1.464  & \quad 1.436 \quad & \quad 2.881 \quad & \quad 4.026 \quad &  \quad 0.68(11) \quad & 0.732 & 0.469\\
$^{140}$Ce & 1.413  & \quad 1.596 \quad & \quad 2.464 \quad & \quad 3.644 \quad &  \quad 0.314(60)  \quad & 0.669 & 0.478\\
$^{142}$Nd & 1.366  & \quad 1.576 \quad & \quad 2.084 \quad & \quad 3.424 \quad &  \quad 0.197(33)  \quad & 0.655 & 0.449\\
$^{144}$Sm & 1.322  & \quad 1.660 \quad & \quad 1.810 \quad & \quad 3.424 \quad &  \quad 0.219(37)  \quad & 0.621 & 0.329\\[0.1cm]
\hline
\hline         
\end{tabular}
\end{table*}

\section{Conclusions}\label{sub::concl}

We have measured the E1 strength distribution up to the neutron
threshold in the semi-magic nucleus $^{136}$Xe using the NRF method
completing a systematic survey on the stable even N=82 isotones to
investigate the PDR. Similar as in the other N=82 isotones, the E1
strength shows a resonance-like concentration well below the neutron
separation energy which shows a strong fragmentation. A detailed
comparison of the experimental results to calculations within the QPM
has been performed for the investigated N=82 isotones. The good
agreement of the calculations in the fragmentation of the E1 strength
confirms that the coupling of a collective mode to complex
configurations used in the model is an accurate description of the
damping of the PDR. Concerning the integrated strength, the influence
of the limited experimental sensitivity has been investigated by
applying the same limit to the QPM results. The agreement of
calculation and experiment is improved by considering the experimental
sensitivity limit. This reveals that part of the experimentally
observed enhancement of the integrated strength for larger
neutron-to-proton ratio is due to the different influences of the
experimental sensitivity limit due to a different amount of
fragmentation along this chain of isotones. But even with this effect
taken into account the enhancement of the strength within the PDR
region with larger N/Z ratio remains which fits into the picture of
an oscillation of a neutron skin as the origin of the low-lying E1
strength.

In this paper, we have presented that the fragmentation of the PDR
observed in high-resolution photon scattering experiments on the one
hand provides in itself an observable to compare to theoretical model
predictions to study the damping mechanism of this mode and on the
other hand has an important impact on the experimental determination
of integral quantities such as the total integrated strength. However,
for a further investigation of the structure of the PDR, complementary
experiments to photon scattering are mandatory. Recent experiments
using the method of $(\alpha ,\alpha' \gamma )$ \cite{savr06a} on Z=50
and N=82 nuclei have revealed a splitting of the low-energy part of
the E1 strength into two groups with different underlying structure
\cite{savr06b, endr09a, endr10}. These unexpected results show that
especially experiments using complementary probes will provide
additional observables to further constrain microscopic model
calculations.

\begin{acknowledgments}

The authors thank the S-DALINAC group around R. Eichhorn for providing
different beam set-ups and continous support during the measurements
and acknowledge the help of M. Babilon, W. Bayer, D. Galaviz Redondo,
T. Hartmann, B. \"Ozel, and S. Volz during the beamtimes. We further
thank P. von Brentano, U. Kneissl, M. Krticka, E. Litvinova, P. von
Neumann-Cosel, A. Richter. A. Tonchev, J. Wambach, and H.J. W\"ortche
for stimulating discussions.  This work was supported by the Deutsche
Forschungsgemeinschaft (contracts SFB 634 and Zi 510/4-1) and by the
Helmholtz Alliance Program of the Helmholtz Association, contract
HA216/EMMI "Extremes of Density and Temperature: Cosmic Matter in the
Laboratory".

% by the LOEWE program of the State of Hesse (Helmholtz
%International Center for FAIR).

\end{acknowledgments}

\bibliographystyle{../../../bibtex/apsrev}
\bibliography{../../../bibtex/english}

\end{document}